\documentclass[aps,prb,amsmath,amssymb,superscriptaddress,twocolumn,10pt]{revtex4-1}

\usepackage{amssymb, amsmath, esint}
\usepackage{graphicx,color}
\usepackage{graphics}
\usepackage{subfigure}
\usepackage{comment}
\usepackage{bbold}

\usepackage[colorlinks]{hyperref}
\hypersetup{
	linkcolor=blue,
}

\usepackage{epsfig}
\usepackage{dcolumn}
\usepackage{bm}

\usepackage{esint}

\usepackage{scrhack}
\usepackage{physics}
\usepackage{eurosym}

\newcommand\ee{\mathrm{e}}

\renewcommand{\ref}[1]{S\ref{#1}}

\def\s{\bm{s}}

\def\H{\mathcal{H}}
\def\PT{\mathcal{PT}}
\def\P{\mathcal{P}}
\def\T{\mathcal{T}}
\def\C{\mathbb{C}}

\begin{document}
	
	\title{Exceptional points in classical spin dynamics}
	\author{Alexey Galda}
	\affiliation{James Franck Institute, University of Chicago, Chicago, Illinois 60637, USA}
	\affiliation{Materials Science Division, Argonne National Laboratory, 9700 S. Cass Avenue, Argonne, Illinois 60439, USA}
	\author{V.\,M. Vinokur}
	\affiliation{Materials Science Division, Argonne National Laboratory, 9700 S. Cass Avenue, Argonne, Illinois 60439, USA} 
	\date{\today}

\begin{abstract}
Non-conservative physical systems admit a special kind of spectral degeneracy, known as exceptional point (EP), at which eigenvalues and eigenvectors of the corresponding non-Hermitian Hamiltonian coalesce. Dynamical parametric encircling of the EP can lead to non-adiabatic evolution associated with a state flip, a sharp transition between the resonant modes. Physical consequences of the dynamical encircling of EPs in open dissipative systems have been explored in optics and photonics. Building on the recent progress in understanding the parity-time ($\PT$)-symmetric dynamics in spin systems, we use topological properties of EPs to implement chiral non-reciprocal transmission of a spin through the material with non-uniform magnetization, like helical magnet. We consider an exemplary system, spin-torque-driven single spin described by the time-dependent non-Hermitian Hamiltonian. We show that encircling individual EPs in a parameter space results in non-reciprocal spin dynamics and find the range of optimal protocol parameters for high-efficiency asymmetric spin filter based on this effect. Our findings offer a platform for non-reciprocal spin devices for spintronics and magnonics.
\end{abstract}

\maketitle

\section*{Introduction}
In magnetic structures, non-reciprocal spin transport  is usually mediated by either surface modes~[\onlinecite{Damon61}] or asymmetric exchange interaction~[\onlinecite{Dzya58}, \onlinecite{Moriya60}], giving rise to asymmetric spin-wave dispersion with respect to the propagation direction~[\onlinecite{Zakeri10, Iguchi15, Sato16}]. Here we present a general method for setting non-reciprocal spin transport based on the non-Hermitian dynamics around exceptional points (EPs), spectral degeneracies arising in non-conservative systems endowed with simultaneous parity-time ($\PT$) invariance~[\onlinecite{Bender98}, \onlinecite{Bender99}]. 
The emerging effect of non-reciprocal transport is topological, hence fundamentally universal, and stems from the adiabaticity breaking unique non-Hermitian systems. The resultant non-reciprocity is highly tunable and robust with respect to noise.

Non-Hermitian extension of Hamiltonian formalism provides a quantitative description of non-conservative dynamics in out-of-equilibrium systems~[\onlinecite{Moiseyev}]. Non-Hermitian Hamiltonians admit a special class of degeneracy points, called exceptional points, where some of their eigenvalues and the corresponding eigenvectors coalesce. This property is unique to non-Hermitian operators and must be distinguished from Hermitian spectrum degeneracies, like diabolical points, where only eigenvalues coalesce. Non-Hermitian degeneracies are stable against both Hermitian or non-Hermitian perturbations, and have far reaching physical implications due to their topological structure that have been explored both theoretically~[\onlinecite{Heiss2000, Heiss2001, Heiss2004, Berry2014}] and experimentally~[\onlinecite{Dembowski2001, Doppler2016, Choi2017}].

Appearance of EPs in the eigenspectrum of non-Hermitian Hamiltonians is a direct consequence of the $\PT$ symmetry-breaking~[\onlinecite{Bender98}, \onlinecite{Bender99}]. Hamiltonians invariant under the simultaneous operations of parity ($\P$) and time-reversal ($\T$) form a special class of non-Hermitian Hamiltonians that admit a fully real eigenspectrum, provided that characteristic amplitudes of their anti-Hermitian parts are below certain threshold values. The $\PT$ symmetry-breaking marks the transition from the real to complex eigenspectrum of $\PT$-symmetric Hamiltonians. The physical interpretation of $\PT$ symmetry as ``balanced gain and loss''~[\onlinecite{Ruschhaupt05}] triggered an increasing number of experimental realizations of $\PT$-symmetric systems and established $\PT$ symmetry-breaking as a non-equilibrium phase transitions between stationary and nonstationary dynamics in driven dissipative systems.

\begin{figure*}[!ht]
	\centering
	\includegraphics[width=2\columnwidth]{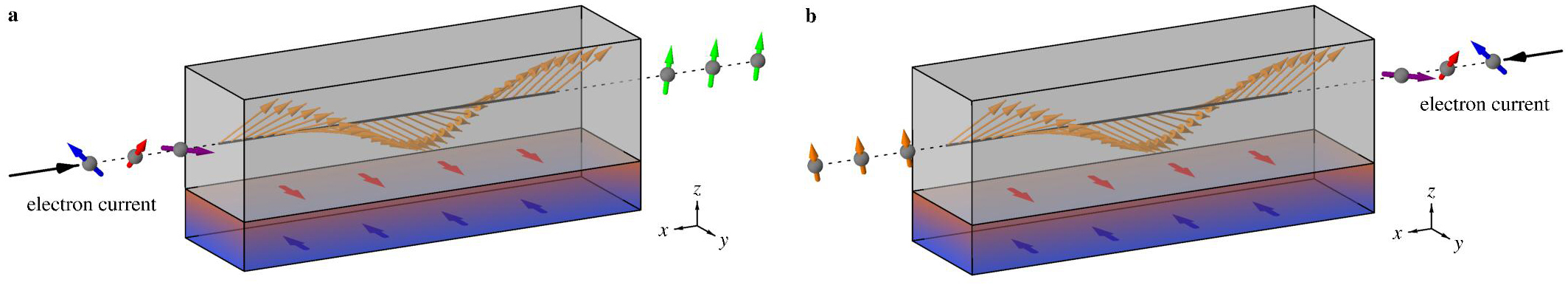}
	\caption{Asymmetric spin-filter effect achieved by sending nonpolarized electrons through a medium with spatially non-uniform magnetization and biased by spin-torque. The combination time-dependent effective magnetic field and spin-torque experienced by individual spins creates a protocol of parametrically encircling an EP of the corresponding effective non-Hermitian Hamiltonian's complex eigenspectrum, resulting in non-reciprocal spin transport properties. Spins sent left-to-right (a) and right-to-left (b) come out polarized in different directions, controlled by the magnitudes of spin-torque and effective magnetic field experienced by the spins. Blue and red flat arrows (anti-)parallel to the $y$-axis represent spin accumulations generated, e.g., by spin Hall effect. As a result, spin-torque is exerted on incident spins propagating through a medium (gray) with spatially varying magnetic field (orange arrows).}
	\label{fig1}
\end{figure*}

We demonstrate that dynamical encircling a second-order EP in the parameter space of a classical magnetic system can lead to non-trivial spin transport properties never seen before. We consider a linear non-Hermitian Hamiltonian describing dynamics of a single classical spin in the external time-dependent magnetic field. This is a generic model for ballistic spin transport experiments, in which spin current is generated by passing non-polarized electrons through a material with non-uniform magnetization, e.g. helical magnet. We show that the combination of applied spin-transfer torque (STT) and time-dependent magnetic field, coordinated in the form of a protocol that takes the system around an EP of the corresponding non-Hermitian Hamiltonian, enables non-reciprocal spin dynamics and transport with a high degree of spin-polarization efficiency.

To derive the propagation of individual spins through a medium with spatially non-uniform magnetic field, we employ the model of a stationary spin in time-dependent magnetic field. We introduce an external imaginary magnetic field to account for the effects of non-conservative forces on spin dynamics, e.g. spin-torque. Indeed, it was shown that the action of Slonczewski STT~[\onlinecite{Slon96}] is equivalent to that of applied imaginary magnetic field in the non-Hermitian Hamiltonian formalism~[\onlinecite{Galda16}]. We propose a design of a classical single-spin system with the inherent $\PT$ symmetry, such that the energy spectrum of the corresponding non-Hermitian Hamiltonian admits a pair of second-order EPs. We then consider a cyclic variation of the applied magnetic field, such that a single EP is encircled in the two-dimensional parameter space $(h_x, h_y)$ of the model, where $h_x$ and $h_y$ are projections of applied magnetic field. Encircling protocols can be constructed in such a way that only a clockwise or anti-clockwise direction, depending on the chirality of the EP, yields the final spin state that is different from the initial one at the beginning of the protocol, despite identical initial and final system parameters. The effect relies on the breakdown of the adiabatic theorem in non-Hermitian systems, and results in non-reciprocal time evolution of spin.

We perform numerical simulations to verify the robustness of the non-reciprocal effect and to identify optimal protocol parameters for achieving maximum asymmetric spin-polarization efficiency in the proposed scheme. In particular, for encircling protocols centered around an isolated EP, we calculate deviations from perfect non-reciprocity as a function of protocol frequency, radius, and initial conditions. Our results show that the highest degree of spin-filter asymmetry between the clockwise and anti-clockwise circular protocols is achieved for slow-frequency parametric trajectories starting in a wide region near the line of $\PT$-symmetry and with a large enough radii to encircle both EPs, as opposed to encircling an individual EP.

Experimental realization of the proposed spin evolution protocols enables a novel type of high-efficiency asymmetric spin-filter devices possessing a number of unique properties. First, high efficiency of the spin-filter effect is guaranteed by the topology of trajectories encircling the EP. Second, the proposed spin-filter device, Fig.~\ref{fig1}, is asymmetric, i.e. breaks time-reversal symmetry. As a result, the polarization directions of spins entering the device from the left and from the right are different. Third, both spin-polarization directions are controlled by the amplitudes of the bias STT and external magnetic field, rendering a highly tunable and efficient system.

\section*{Results}

The most general form of a linear spin Hamiltonian contains three independent parameters~[\onlinecite{Galda17}], which, without any loss of generality, can be chosen as two components of the real magnetic field along the $x$ and $y$ axes, and a single component of the imaginary field along the $y$ axis:
\begin{equation}
\H_0 = h_x S_x + (h_y + i \beta)\,S_y\,,\label{H0}
\end{equation}
where $\beta$ is the dimensionless amplitude of applied spin-torque~[\onlinecite{Galda16}].

\begin{figure*}[!ht]
	\centering
	\includegraphics[width=2\columnwidth]{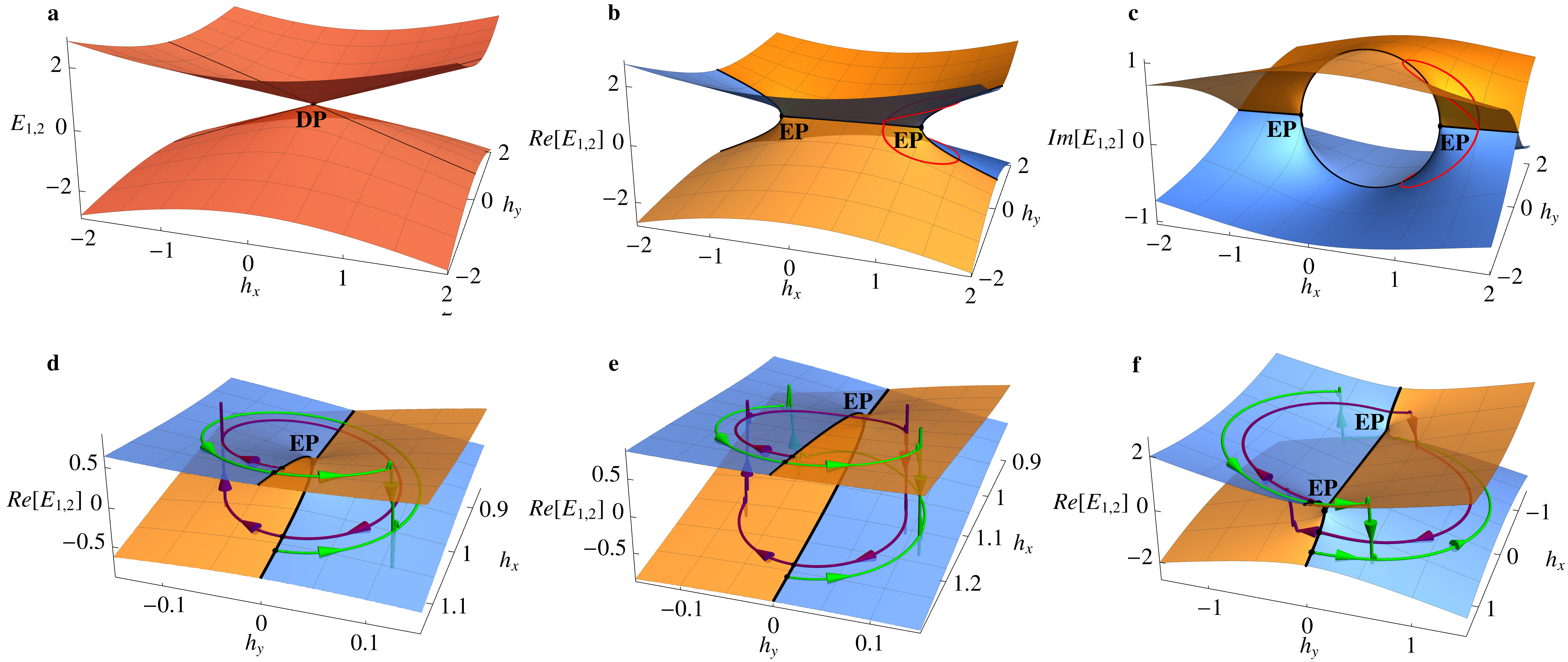}
	\caption{(a) Diabolical point (DP) of real energy spectrum of the Hermitian Hamiltonian~(\ref{H1}) with $\beta = 0$. DP bifurcates into two EPs, shown in (b), (c) for $\beta = 1$ in the real and imaginary parts of eigenspectrum, correspondingly. Na\"{\i}ve consideration predicts that a $4\pi$-rotation (red trajectory) around an EP would bring the system back to its initial state (up to a phase factor). (d)--(f) Numerical simulations of state evolution during clockwise (purple) and anti-clockwise (green) encircling of the EP, illustrating non-adiabatic state-flip events as sharp jumps between two spectral surfaces. The system's final state after a complete $2\pi$-rotation depends on the encircling direction, illustrating non-reciprocity of non-Hermitian parametric driving. The non-reciprocal time evolution is shown for the cases of (d) encircling of a single EP, (e) circular trajectory in the vicinity of the EP, and (f) complete encircling of both EPs.}
	\label{fig2}
\end{figure*}

To describe single-spin dynamics governed by the non-Hermitian Hamiltonian~(\ref{H0}), we employ SU(2) spin-coherent states~[\onlinecite{Lieb, Garg, Radcliffe71}]: ${\ket{z} = \ee^{z S_+}\ket{S, -S}}$, where ${S_\pm \equiv S_x \pm i S_y}$, and ${z \in \C}$ is the standard stereographic projection of the spin direction on a unit sphere, ${z = (s_x + is_y)/(1 - s_z)}$, with ${s_i \equiv S_i/S}$.
For the purpose of this work, we focus on the classical limit of large spin, $S \to \infty$, in which case one can write the generalized spin equation of motion as follows~[\onlinecite{Galda16}]:
\begin{equation}\label{zHamilton}
\dot z = i\frac{ \left(1 + \bar z z\right)^2}{2S} \frac{\partial \H_0 (z, \bar z)}{\partial \bar z}\,,
\end{equation}
where $\H_0 (z, \bar z)$ is the expectation value in spin-coherent states:
\begin{equation}
\H_0 (z, \bar z) = \frac{\expval{\H_0}{z}}{\bra{z}\ket{z}}\,.\label{Hexp}
\end{equation}
For the spin Hamiltonian~(\ref{H0}), this yields the following evolution equation:
\begin{equation}\label{zdot}
\dot z(t) = -i\,\frac{\,h_x - i(h_y + i\beta)}2\!\left[ z^2 - \frac{h_x + i(h_y + i\beta)}{h_x - i(h_y + i\beta)}\right]\,.
\end{equation}
The eigenvalue problem for the spin Hamiltonian~(\ref{H0}) in the classical limit reduces to evaluating the expectation values, Eqs.~(\ref{Hexp}), at the fixed points of the corresponding classical spin dynamics on a unit sphere, Eq.~(\ref{zHamilton})--(\ref{zdot}). For the linear non-Hermitian Hamiltonian~(\ref{H0}), this is also equivalent to the eigenvalue problem for a two-level system described by the Hamiltonian:
\begin{equation}
\H = h_x \sigma_x + (h_y + i \beta) \sigma_y\,,\label{H1}
\end{equation}
where $\sigma_i$ are Pauli matrices, which yields the following complex eigenspectrum:
\begin{equation}
E_{1, 2} = \pm\sqrt{h_x^2 + ( h_y + i\beta)^2}\,.\label{E12}
\end{equation}
The two fixed points of classical spin dynamics governed by the Hamiltonian~(\ref{H0}), referred to as $A$ and $B$ in the text, are defined by the following expression in stereographic projection coordinates:
\begin{equation}
z_{A,B} = \pm\sqrt{\frac{h_x + i(h_y + i \beta)}{h_x - i(h_y + i \beta)}}\,.\label{zAB}
\end{equation}

To derive non-reciprocal time evolution of classical spin, we consider the eigenspectrum of the Hamiltonian~(\ref{H1}) in two-dimensional parameter space of mutually orthogonal magnetic fields, $(h_x, h_y)$, at fixed $\beta$. Figs.~\ref{fig2}(a)--\ref{fig2}(c) show the real and imaginary parts of the eigenspectrum~(\ref{E12}) at (a) $\beta = 0$ and (b), (c) $\beta = 1$. In the first (Hermitian) case, the spectral surfaces meet at a single diabolical point (DP) $h_x = h_y = \beta = 0$. It is a spectral degeneracy that occurs due to simple symmetries and disappears under generic perturbations. In the Hermitian case, states on both spectral surfaces are stable, with two fixed points of classical spin dynamics classified as centers. Distinct surface colors in Figs.~\ref{fig2}(b)--\ref{fig2}(f) correspond to opposite signs of imaginary parts of the eigenvalues, with the states on the blue (orange) energy surface being dynamically (un)stable, i.e. corresponding to spin directions of (un)stable equilibria. At $\beta \neq 0$, i.e. in the non-Hermitian case, the diabolical point splits into two chiral EPs located at $(h_x = \pm\beta, h_y = 0)$. The black solid line $h_y = 0$ in Figs.~\ref{fig2}(b)--\ref{fig2}(f) is the line of $\PT$ symmetry, where the Hamiltonian~(\ref{H0}) is $\PT$-symmetric~[\onlinecite{Galda16}, \onlinecite{Galda17}]. The interval of this line between the EPs corresponds to the regime of broken $\PT$ symmetry, with the eigenvalues forming complex conjugate pairs, while on the part of the line outside of this interval the eigenvalues are purely real. Note that in the general case, $h_y \neq 0$, the Hamiltonian~(\ref{H0}) is not $\PT$-symmetric and has a complex eignespectrum. It is the reducibility of the non-Hermitian Hamiltonian to the $\PT$-symmetric form that produces EPs in the eigenspectrum.

Figs.~\ref{fig2}(b), \ref{fig2}(c) show the na\"{\i}ve picture of encircling a single EP in 2-dimensional parameter space, with the state evolution following a $4\pi$-periodic closed trajectory (red line) by visiting both, top and bottom, spectral surfaces before returning to the starting point. This picture is generally incorrect in the case of dynamic encircling of an EP with control parameters varied cyclically at finite speed. Instead of adiabatic evolution, the system experiences sudden jumps between the two spectral surfaces, in the direction from the unstable to the stable one, see Figs.~\ref{fig2}(d)--\ref{fig2}(f). The jumps occur not immediately upon crossing the line of $\PT$ symmetry into the unstable region, but after a certain time interval, illustrating the effect of stability loss delay~[\onlinecite{Doppler2016}].

Let us consider the physics of encircling an isolated EP in parameter space $(h_x,\, h_y)$ of the model~(\ref{H0}) by applying time-dependent magnetic field. Encircling an EP reveals its chiral nature and provides access to topological non-reciprocal physical effects unique to non-Hermitian systems. In Fig.~\ref{fig2}(d), we present parametric trajectories starting on different energy surfaces and going around the EP in opposite directions. For fixed model parameters $(h_x,\, h_y, \beta)$, the $z$-component of the starting point of each trajectory, i.e. the value of complex energy at $t = 0$, is determined by the initial spin orientations $\s(0)$, which can take any direction on the unit sphere. For the purpose of Figs.~\ref{fig2}(b)--\ref{fig2}(f), we only consider the initial spin orientations corresponding to the two fixed points of spin dynamics A and B, which confines the start and end points of the trajectories in Figs.~\ref{fig2}(b)--\ref{fig2}(f) to the spectral surfaces and best illustrates non-reciprocal dynamics. Numerical simulations of spin evolution show that the end states for parametric trajectories starting on the top and bottom spectral surfaces in Fig.~\ref{fig2}(d) and encircling the EP in closed loops of the same orientation always coincide. On the contrary, protocols with opposite encircling directions lead to mutually different end states, either on the top or bottom energy surfaces, and, therefore, different final spin orientations after any number of complete revolutions around the EP. It has been pointed out in the literature that in order to observe such non-reciprocal behaviour, the start/end points of closed loop trajectories much lie on the $\PT$-symmetry line, where both eigenstates are neither stable nor unstable~[\onlinecite{Uzdin11, Moiseyev2013, Rabl2015, Doppler2016, Christo2017, Choi2017, Zhang2018}]. In fact, the effect of non-reciprocal dynamics stems from the process of crossing the line of $\PT$-symmetry in opposite directions and not strictly from the topology of parametric trajectories. It turns out that to achieve non-reciprocity it is sufficient to realize a closed loop protocol in the vicinity of a single EP without actually encircling it~[\onlinecite{Hassan2017}], see Fig.~\ref{fig2}(e), or to encircle both EPs, as shown in Fig.~\ref{fig2}(f).

Consider spin dynamics governed by the Hamiltonian~(\ref{H1}) under the following two protocols of encircling the isolated EP $(\beta = 1,\, h_x = h_0,\, h_y = 0)$:
\begin{align}
&P_{1,2}(t) \!=\! \left[h_x(t),\, h_y(t) \right] \!=\! \left[h_0 + R_0\cos(\phi(t)),\,R_0\sin(\phi(t))\right]\,,\nonumber\\
&\phi(t) =\phi_0 \mp \omega t\,. \label{P12}
&\end{align}
For the purpose of this work, we focus on circular trajectories centered at the EP, with $h_0 = 1$, see Fig.~\ref{fig2}(d). Suppose the system is initialized at $t = 0$ with phase $\phi_0 = 0$ and subjected to the protocol $P_1$, which takes the system clockwise around the EP. Regardless of the initial spin orientation, i.e. location of the starting point with respect to the two spectral surfaces, the spin will eventually saturate in the direction of the stable fixed point and quickly converge to the state on the top energy surface, even if it was initialized in the unstable fixed point (in which case the transition occurs rapidly via a non-adiabatic state-flip process after a finite time delay). The clockwise protocol $P_1$ ensures that the system's final state after a full rotation around the EP is on the bottom spectral surface, as shown in Fig.~\ref{fig2}(d). Following a completely analogous argument, one can see that the encircling protocol $P_2$, which drives the system anti-clockwise around the EP, results in a different final state - on the top spectral surface, regardless of the initial spin direction at $t = 0$. This illustrates the chiral nature and non-reciprocity of parametrically driven dynamics near the EP.

\begin{figure*}[!htb]
	\centering
	\includegraphics[width=2\columnwidth]{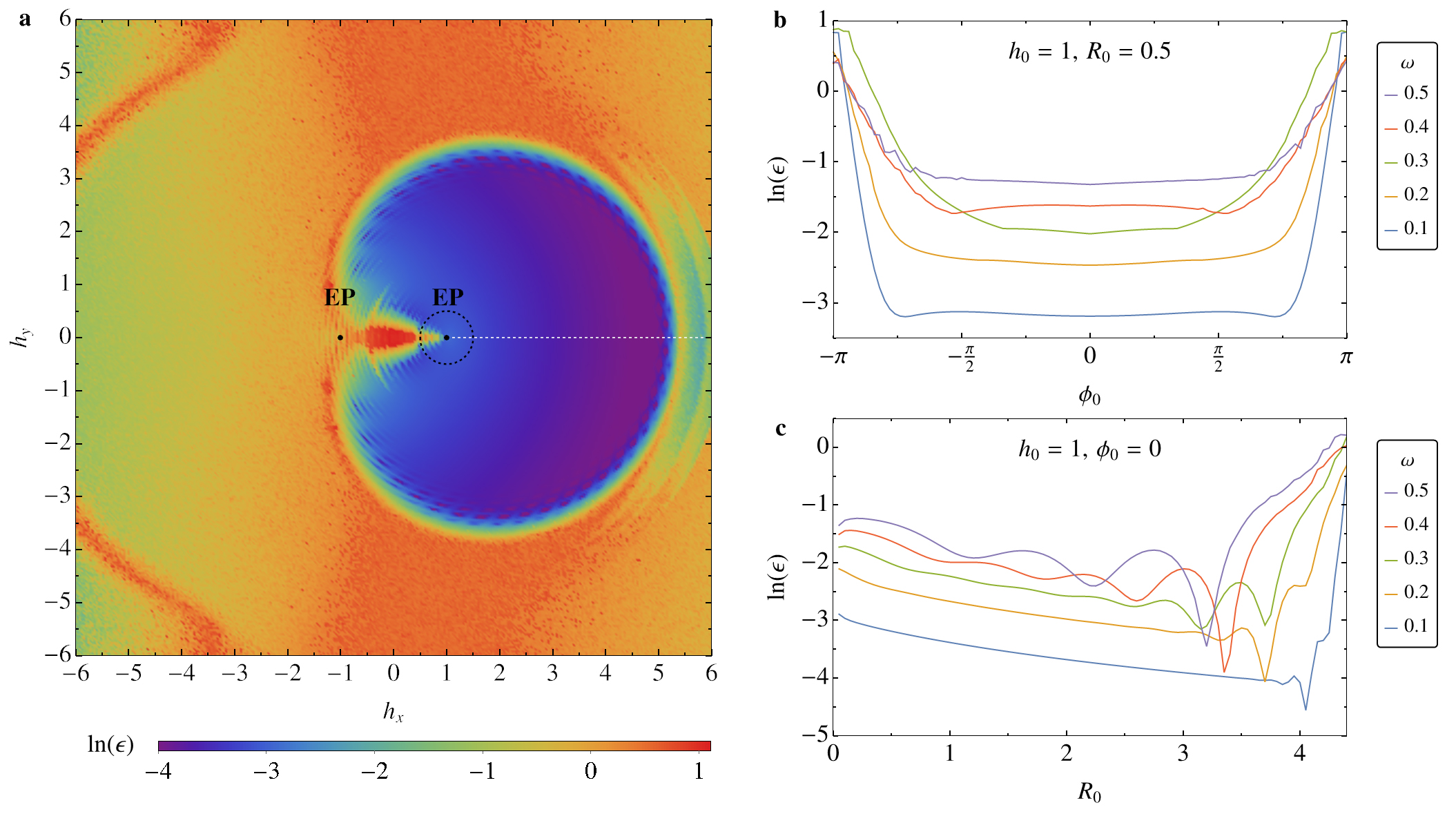}
	\caption{(a) Color map of maximum deviation parameter, $\epsilon$, of the final spin direction from the fixed points $z_{A,B}$ after encircling protocols $P_{1,2}$, as a function of initial protocol phase $\phi_0$ (deviation from the line of $\PT$-symmetry, $h_y = 0$) and protocol frequency $\omega$, plotted in logarithmic scale. The deviation $\epsilon$ is a measure of effectiveness of the non-reciprocal spin-filter effect. (b) Dependence of the deviation parameter on the initial phase for a range of protocol frequencies, showing that slower encircling protocols yield higher non-reciprocal efficiency. The plot corresponds to encircling protocols starting and ending on the black dashed line in panel (a). (c) The deviation parameter as a function of protocol radius $R_0$. A sharp jump in deviation from perfect non-reciprocity is observed for large radii within the considered range, corresponding to the boundary of the high-efficiency region in panel (a), shown in blue. The size of the region with small deviation parameter $\epsilon$ shrinks with increasing frequency. The plot corresponds to encircling protocols starting and ending on the white dashed line in panel (a).
	}
	\label{fig3}
\end{figure*}

As we have pointed out, in order to exploit the non-reciprocal properties discussed above, the encircling protocols must start and end near the line of $\PT$ symmetry. Only such trajectories in parameter space can take the system to different final states upon crossing the line of $\PT$ symmetry in opposite directions, e.g. by completing a single clockwise or anti-clockwise closed loop around the EP. The question of tolerance to deviations from this condition, as well as optimization of other protocol parameters to achieve consistent non-reciprocal behaviour, have not been thoroughly studied in the literature. Here we investigate the range of initial conditions and parameters of the encircling protocols $P_{1,2}$ optimal for observing the non-reciprocal evolution. We performed numerical simulations of spin dynamics for a range of initial phases $\phi_0$, radii $R_0$, and driving frequencies $\omega$ of the encircling protocols, Eq.~(\ref{P12}). The final state of the system after executing the protocols $P_{1,2}$ correspond to the fixed points $z_{A,B}$, see Eq.~(\ref{zAB}). To quantify deviation of the end spin state after each encircling protocol from the corresponding fixed point, we define deviation angles $\epsilon_{i}(\phi_0, \omega)$ between the spin direction at the end of a protocol $P_i$, $i = 1,2$, and the direction corresponding to fixed point $z_{A,B}$. As an absolute measure of deviation from perfect non-reciprocity, we take the maximum deviation between the two protocols, $\epsilon(\phi_0, \omega) \equiv \max[\epsilon_{1}(\phi_0, \omega), \epsilon_{2}(\phi_0, \omega)]$.

In Fig.~\ref{fig3}(a) we present the results of numerical simulations of spin dynamics following the protocols $P_{1,2}$, Eq.~(\ref{P12}), with $h_0 = 1$ and $\omega = 0.1$, centered around the EP, for a range of parameters $R_0$ and $\phi_0$, which determine the starting point of both protocols in the $(h_x,\, h_y)$-plane. The color map shows the maximum deviation parameter $\epsilon$, measured in radians, in the logarithmic scale. A wide range of initial protocol parameters $(h_x, h_y)$ that favor high-efficiency non-reciprocal spin-filter effect lie in the central blue region around the EP. The interval directly between the EPs exhibits poor non-reciprocity because it is adjacent to the region of broken $\PT$-symmetry of the Hamiltonian~(\ref{H0}) with $h_y = 0$, where the eigenspectrum is purely imaginary, see Fig.~\ref{fig2}(b)--\ref{fig2}(c), corresponding to one stable and one unstable fixed points of spin dynamics. As a result, the final states of the system in the vicinity of this region in the $(h_x, h_y)$-plane are almost identical regardless of the protocol encircling direction. The boundary of the low deviation region is set by specific properties of the stability loss delay effect, discussing which is beyond the scope of the present work and will be addressed elsewhere. Note that the highest-efficiency non-reciprocal protocols in Fig.~\ref{fig2}(a) have a relatively large radius, such that they encircle both EPs. This is an important result that extends the idea of encircling a single EP to achieve non-reciprocity to parametric trajectories that go around both isolated EPs to achieve an even more robust effect.

In Fig.~\ref{fig3}(b), the deviation quantity $\epsilon$ is plotted for protocols starting on the circle of radius $R_0 = 0.5$ around the EP, shown by the black dashed line in Fig.~\ref{fig3}(a), for a range of initial phases $\phi_0$ and protocol frequencies $\omega$. This result illustrates that lower encircling frequencies result in smaller deviation parameter $\epsilon$, i.e. higher efficiency of the non-reciprocal effect. Fig.~\ref{fig3}(c) shows dependence of the deviation parameter on protocol radius $R_0$ for the same range of frequencies $\omega$, calculated at fixed initial phase $\phi_0 = 0$. The initial conditions for the protocols studied in Fig~\ref{fig3}(c) are chosen on the line of $\PT$-symmetry, shown as a white dashed line in Fig.~\ref{fig3}(a). In the limit of slow encircling speeds, the protocols with larger radius result in smaller deviation $\epsilon$, even when both EPs are encircled by the protocols ($R_0 > 2$). The example of non-reciprocal spin dynamics that results from parametric encircling of both EPs is presented in Fig.~\ref{fig2}(f). While Fig.~\ref{fig3}(c) shows relatively low degree of deviation from non-reciprocity in the case of parametric trajectories with small radius $R_0$, the practical importance of protocols that lie in the immediate vicinity of the EP is weakened by the fact that eigenstates coalesce in the EP. As a result, the angle between spin orientation given by the fixed points $z_A$ and $z_B$ disappears in the limit $R_0 \to 0$.

Note that it is not necessary to encircle an EP to observe non-reciprocity, see Fig.~\ref{fig2}(e). Periodic trajectories in the vicinity of an EP can achieve the same effect~[\onlinecite{Hassan2017}]. What is crucial for realizing non-reciprocal dynamics is the direction in which the protocols cross the line of $\PT$-symmetry, where fixed points of the corresponding dynamics change their stability properties. In fact, parametric trajectories may not enclose any area at all and still achieve a high-efficiency non-reciprocal state conversion. The study of various trajectory types and the robustness of resultant non-reciprocal dynamics will be considered elsewhere.

\section*{Methods}
The numerical simulations were carried out in Wolfram Mathematica~[\onlinecite{Mathematica}]. Each data point in Figs.~\ref{fig3}(a)--\ref{fig3}(d) was obtained by averaging results of 10 separate simulation runs with arbitrary initial spin orientations uniformly distributed on a unit sphere at protocol initiation.

\section*{Discussion}
To summarize, there has been a remarkable advance in exploiting non-Hermitian systems and their distinct feature, spectral exceptional points. The stage for this non-Hermitian physics is however restricted mostly to optical and nanophotonic systems.  Here we extended the exploration of EP singularities onto another broadest class of physical systems, spin systems. We demonstrated that spin systems can host EP singularities and thus become a crucial element for realizing devices exhibiting non-trivial topological transport seen so far only in optical systems. This establishes that non-reciprocal effects are ubiquitous in open dissipative systems with gain and loss. Implementation of topological spin transfer enabled us to propose a high-efficiency asymmetric spin filter device based on non-reciprocal spin dynamics around an EP of the corresponding non-Hermitian Hamiltonian. As a prototypical model, we studied time evolution of a single classical spin in presence of time-dependent magnetic field and spin-transfer torque. Appearance of EPs in the Hamiltonian's eigenspectrum allows for encircling protocols that result in different final spin states despite identical initial conditions. We considered two protocols with opposite encircling orientations and showed that such a scheme can be used as a high-efficiency spin filter. Optimal protocol parameters are suggested, indicating that encircling trajectories in parameter space of applied magnetic fields must start and end near the line of $\PT$ symmetry, with slower encircling frequencies resulting in higher efficiency of the asymmetric spin-filter effect.

\section*{Acknowledgements}

A.G. and V.M.V. were supported by the U.S. Department of Energy, Office of Science, Basic Energy Sciences, Materials Sciences and Engineering Division.


\begin{thebibliography}{1}
	
	\bibitem{Damon61}
	Damon,\,R.\,W. \& Eshbach,\,J.\,R. Magnetostatic modes of a ferromagnet slab. \textit{J. Phys. Chem. Solids} \textbf{19}, 308-320 (1961).
	
	\bibitem{Dzya58}
	Dzyaloshinsky,\,I.\,E. A thermodynamic theory of ``weak'' ferromagnetism of antiferromagnetics. \textit{J. Phys. Chem. Solids} \textbf{4}, 241--255 (1958).
	
	\bibitem{Moriya60}
	Moriya,\,T. Anisotropic superexchange interaction and weak ferromagnetism. \textit{Phys. Rev.} \textbf{120}, 91 (1960).
	
	\bibitem{Zakeri10}
	Zakeri, K. \textit{et al.} Asymmetric spin-wave dispersion on Fe(110): direct evidence of the Dzyaloshinskii-Moriya interaction. \textit{Phys. Rev. Lett.} \textbf{104}, 137203 (2010).
	
	\bibitem{Iguchi15}
	Iguchi,\,Y., Uemura,\,S., Ueno,\,K \& Onose,\,Y., Nonreciprocal magnon propagation in a noncentrosymmetric ferromagnet ${\text{LiFe}}_{5}{\text{O}}_{8}$. \textit{Phys. Rev. B.} \textbf{92}, 184419 (2015).
	
	\bibitem{Sato16}
	Sato, T.\,J.\,\textit{et al.} Magnon dispersion shift in the induced ferromagnetic phase of noncentrosymmetric MnSi, \textit{Phys. Rev. B} \textbf{94}, 144420 (2016).
	
	\bibitem{Bender98} 
	Bender,\,C.\,M. \& Boettcher,\,S. Real spectra in non-Hermitian Hamiltonians having $\PT$ symmetry. \textit{Phys. Rev. Lett.} \textbf{80}, 5243 (1998).
	
	\bibitem{Bender99}
	Bender,\,C.\,M. Boettcher,\,S. \& Meisinger, P.\,N. $\PT$-symmetric quantum mechanics. \textit{J. Math. Phys.} \textbf{40}, 2201--2229 (1999).
	
	\bibitem{Moiseyev}
	Moiseyev,\,N. \textit{Non-Hermitian quantum mechanics.} (Cambridge Univ. Press, 2011).
	
	\bibitem{Heiss2000}
	Heiss,\,W.\,D. Repulsion of resonance states and exceptional points. \textit{Phys. Rev. E} \textbf{61}, 929 (2000).
	
	\bibitem{Heiss2001}
	Heiss,\,W.\,D. \& Harney,\,H.\,L. The chirality of exceptional points. \textit{Eur. Phys. J. D} \textbf{17}, 149--151 (2001).
	
	\bibitem{Heiss2004}
	Heiss,\,W.\,D. Exceptional points - their universal occurrence and their physical significance. \textit{Czech. J. Phys.}, \textbf{54}, 1091--1099 (2004).
	
	\bibitem{Berry2014}
	Berry,\,M.\,V. Physics of nonhermitian degeneracies. \textit{Czech. J. Phys.} \textbf{54}, 1039--1047 (2004).
	
	\bibitem{Dembowski2001}
	Dembowski,\,C. et al. Experimental observation of the topological structure of exceptional points. \textit{Phys. Rev. Lett.} \textbf{86}, 787 (2001).
	
	\bibitem{Doppler2016}
	Doppler,\,J. \textit{et al.} Dynamically encircling an exceptional point for asymmetric mode switching. \textit{Nature} \textbf{537}, 7618 (2016).
	
	\bibitem{Choi2017}
	Choi,\,Y. \textit{et al.} Extremely broadband, on-chip optical nonreciprocity enabled by mimicking nonlinear anti-adiabatic quantum jumps near exceptional points. \textit{Nat. Commun.} \textbf{8}, 14154 (2017).
	
	\bibitem{Ruschhaupt05}
	Ruschhaupt, A., Delgado, F. \& Muga, J.\,G. Physical realization of $\PT$-symmetric potential scattering in a planar slab waveguide. \textit{J. Phys. A: Math. Gen.} \textbf{38}, L171 (2005).
		
	\bibitem{Slon96}
	Slonczewski, J.\, C. Current-driven excitation of magnetic multilayers.\textit{ J. Magn. Magn. Mater.} \textbf{159}, L1–L7 (1996).
	
	\bibitem{Galda17}
	Galda,\,A. \& Vinokur,\,V.\,M. Linear dynamics of classical spin as M{\"o}bius transformation. \textit{Sci. reports} \textbf{7}, 1168 (2017).
	
	\bibitem{Galda16}
	Galda,\,A. \& Vinokur,\,V.\,M. Parity-time symmetry breaking in magnetic systems. \textit{Phys. Rev. B} \textbf{94}, 020408 (2016).
	
	\bibitem{Lieb}
	Lieb,\,E.\,H. The classical limit of quantum spin systems. \textit{Comm. Math. Phys.} \textbf{34}, 327--340 (1973).
	
	\bibitem{Garg}
	Stone, M., Park, K.-S. \& Garg, A. The semiclassical propagator for spin coherent states. \textit{Journ. Math. Phys.} \textbf{41}, 8025--8049 (2000).
	
	\bibitem{Radcliffe71}
	Radcliffe, J.\,M. Some properties of coherent spin states. \textit{J. Phys. A: Gen. Phys.} \textbf{4}, 313 (1971).
	
	\bibitem{Uzdin11}
	Uzdin, R., Mailybaev, A. \& Moiseyev, N. On the observability and asymmetry of adiabatic state flips generated by exceptional points. \textit{J. Phys. A: Math. Theor.} \textbf{44}, 435302 (2011).
	
	\bibitem{Moiseyev2013}
	Gilary,\,I., Mailybaev,\,A.\,A. \& Moiseyev,\,N. Time-asymmetric quantum-state-exchange mechanism. \textit{Phys. Rev. A} \textbf{88}, 010102 (2013).	
	
	\bibitem{Rabl2015}
	Milburn, T.\,J. \textit{et al.} General description of quasiadiabatic dynamical phenomena near exceptional points. \textit{Phys. Rev. A} \textbf{92}, 052124 (2005).
	
	\bibitem{Christo2017}
	Hassan, A.\,U. \textit{et al.} Dynamically encircling exceptional 	points: exact evolution and polarization state conversion. \textit{Phys. Rev. Lett.} \textbf{118}, 093002 (2017).
	
	\bibitem{Zhang2018}
	Zhang,\,X.\,L., Wang,\,S., Hou,\,B. \& Chan,\,C.\,T. Dynamically encircling exceptional points: in situ control of encircling loops and the role of the starting point. \textit{Phys. Rev. X} \textbf{8}, 021066 (2018).

	\bibitem{Hassan2017}
	Hassan, A.\,U. \textit{et al.} Chiral state conversion without encircling an exceptional point. \textit{Phys. Rev. A} \textbf{96}, 052129 (2017).
	
	\bibitem{Mathematica}
	Inc., W. R. Mathematica, Version 11.0. Champaign, IL, 2016.
	
\end{thebibliography}
\end{document}